# Applying Adapted Big Five Teamwork Theory to Agile Software Development

**Diane Strode**
Faculty of Business and Information Technology
Whitireia Polytechnic
Wellington, New Zealand
Email: diane.strode@alumni.unimelb.edu.au

## Abstract

Teamwork is a central tenet of agile software development but various teamwork theories only partially explain teamwork in that context. Big Five teamwork theory is one of the most influential teamwork theories, but prior research shows that the team leadership concept in this theory it is not applicable to agile software development.  This paper applies an adapted form of Big Five teamwork theory to cases of agile software development. Three independent cases were drawn from a single organisation providing a measure of control over contextual factors affecting the teamwork. The findings show that the adapted form of Big Five teamwork theory, including a shared team leadership concept, is fully applicable to some forms of agile software development, but not all. For practitioners, insights as to which agile practices support teamwork are provided.

## 1    Introduction

Recent decades have been called a golden age for team research (Salas et al. 2008) with organisational teamwork theory arising from various contexts where groups work on critical goal-directed tasks. These teamwork theories are grounded in empirical data from hospital operating theatres, warships, natural disaster emergencies, accident and emergency centres, business strategizing, and distributed virtual work environments (Hollenbeck et al. 2012). Even though agile software development is a mainstream approach to system development and a central tenet of the agile philosophy is the need for effective teamwork (Chow and Cao 2008), this significant body of teamwork research has seldom been applied to the agile software development project context.

The agile approach is adopted in 50% of software development projects world-wide (Stavru 2014), and Gregory et al. (2015) have identified that teamwork in agile projects is of particular concern to practitioners. However, there are conflicting perspectives on teamwork in the agile context. Some empirical research shows that adopting agile practices leads to effective teamwork. For example, Tessem (2014) found that certain agile practices in Scrum teams support team empowerment, which is an important enabler of self-organising agile software development teams. Whitworth and Biddle (2007) found that agile software development enhances and supports teamwork in a study of teams using eXtreme Programming. Similarly, Wood et al. (2013) found that team cooperation depends on agile practices such as collective code ownership and following coding standards. Another perspective is that adopting agile practices is necessary but not sufficient for effective teamwork. Additional supporting practices are needed, such as teamwork training (Gandomani et al. 2015; Vijayasarathy and Turk 2012), team selection (da Silva et al. 2013), and management support for teamwork (Moe et al. 2010). Yet another, conflicting, perspective is proposed by Chan and Thong (2009). They theorise that effective teamwork improves the acceptance and adoption of agile methodologies and argue that "Good teamwork helps reduce the distance, either physically or psychologically, between people, and thus creates an environment that facilitates learning among team members. Developers who have good teamwork are more likely to learn from each other and share knowledge required for agile development." (Chan and Thong 2009, p. 810).

To untangle these perspectives, it is first necessary to have a clear idea of those teamwork theories that are applicable to agile software development, or if we need new theory (Moe and Dingsøyr 2008). Once we have settled on a pertinent teamwork theory, then researchers can address questions about the causal relationships between teamwork, agile adoption, and agile practices. One high-impact and well-grounded teamwork theory, called the Big Five teamwork theory (Salas et al. 2005), has been applied by Moe and Dingsøyr (2008) in a single case study of agile software development. They found that Big Five teamwork theory does apply to agile projects, with one exception. To extend that research, this paper addresses the research question: How does an adapted Big Five teamwork theory fit agile software development project teams?

To address this question, first an adapted Big Five teamwork theory was developed and then its concepts applied to three cases of agile software development in a single organisation. Findings show the adapted theory fits some, but not all types of agile software development project.

The paper is organised as follows. Big Five teamwork theory is described followed by research on teamwork in the context of agile software development. The case study research design and how qualitative content analysis was used for data analysis is described. The background to each case is described, and findings are presented that compare the cases and their fit with the concepts of the adapted Big Five teamwork theory. Contributions to theory and practice are described, and future research directions are set out.

## 1.1 Big Five Teamwork Theory

Work groups and teams are defined by Kozlowski and Bell (2003, p. 6). Teams are "(a) composed of two or more individuals, (b) who exist to perform organizationally relevant tasks, (c) share one or more common goals, (d) interact socially, (e) exhibit task interdependencies (i.e., workflow, goals, outcomes), (f) maintain and manage boundaries, and (g) are embedded in an organizational context that sets boundaries, constrains the team, and influences exchanges with other units in the broader entity". Hollenbeck et al. (2012) identified 42 different types of team in the literature, and more than 138 models and frameworks explain team performance (Salas et al. 2007). This richness makes it difficult to select an appropriate teamwork theory to apply in a study.

In this study, the Big Five theory developed by Salas et al. (2005) was selected and adapted as an appropriate framework for exploring teamwork in cases of agile software development because of its research impact, founding literature, and prior use in the agile context. As at 21 July 2015 Google Scholar showed 715 citations for the Big Five theory, and Scopus showed 351 citations. The theory is well-founded because it is based on a literature review covering 20 years of research involving the analysis of 121 references from multiple contexts. In addition, Moe and Dingsøyr (2008) have assessed the applicability of this theory to agile software development in a single case of Scrum, and found it was broadly applicable to teamwork in that context, with the exception of the way the leadership is enacted (this is discussed later in this paper). The Big Five theory has five components and three coordinating mechanisms, with 10 propositions explaining how they are related. The components and their coordinating mechanisms are defined by Salas et al. (2005) as follows.

*Adaptability* - Ability to adjust strategies based on information gathered from the environment through the use of backup behaviour and reallocation of intra-team resources. Altering a course of action or team repertoire in response to changing conditions (internal or external).

*Backup behaviour* - Ability to anticipate other team members' needs through accurate knowledge about their responsibilities. This includes the ability to shift workload among members to achieve balance during high periods of workload or pressure.

*Closed-loop communication* - The exchange of information between a sender and a receiver irrespective of the medium.

*Mutual performance monitoring* - The ability to develop common understandings of the team environment and apply appropriate task strategies to accurately monitor teammate performance.

*Mutual trust* - The shared belief that team members will perform their roles and protect the interests of their teammates.

*Shared mental models* - An organizing knowledge structure of the relationships among the task the team is engaged in and how the team members will interact.

*Team leadership* - Ability to direct and coordinate the activities of other team members, assess team performance, assign tasks, develop team knowledge, skills, and abilities, motivate team members, plan and organize, and establish a positive atmosphere.

*Team orientation* - Propensity to take other's behaviour into account during group interaction and the belief in the importance of team goal's over individual members' goals.

## 1.2 Teamwork Research in the Agile Context

Agile software development places particular value on teamwork in self-organising teams. Self-organising teams are a critical factor in the success of agile projects (Chow and Cao 2008). A self-organising team is also referred to as self-managing, empowered, or autonomous (Moe et al. 2008) and is defined in management literature as follows. "Self-organization refers to autonomous decision making within a unit with respect to both the transactions (output) it wants to realize and the way it organizes its transformation processes to achieve these transactions" (Balkema and Molleman 1999, p. 135). In systems development, this means a self-organising team makes its own decisions about how to create a requested work product, and the process they will follow. In agile projects, teams are expected to manage their own workload, allocate work among themselves, and participate in shared decision making (Moe et al. 2012; Moe et al. 2008).

Three studies could be located that focus exclusively on applying teamwork theory to agile software development. The first study, by Moe and Dingsøyr (2008), was a single case study of a Scrum project designed to test the applicability of the Big Five theory of teamwork (Salas et al. 2005). They found that each component and coordination mechanism in the theory was reflected in the agile project with the exception of team leadership. The definition of the team leadership role in the Big Five theory is at odds with the accepted leadership profile in an agile project, which promotes shared distributed leadership among all team members. Salas et al. (2005, p. 560) define team leadership and team leadership behaviours as follows "[Team leaders] facilitate team problem solving. Provide performance expectations and acceptable interaction patterns. Synchronize and combine individual team member contributions. Seek and evaluate information that affects team functioning. Clarify team member roles. Engage in preparatory meetings and feedback sessions with the team." In their Scrum study, Moe and Dingsøyr (2008) identified two team roles that involve aspects of team leadership. The Scrum master acting as a coach, facilitates team-based problem solving and "focus[s] on protecting the team against external noise, removing impediments and facilitating the different processes defined by Scrum" (Moe and Dingsøyr 2008, p. 16). The Product owner in a Scrum team participates in preparatory meetings to discuss requirements, which is another leadership behaviour proposed by Salas et al. (2005, p. 560). However, neither of these roles exactly match the team leadership role of Big Five teamwork theory. In the Big Five theory, team leadership resides in a single team leader whereas in an agile software development team there is no one designated team leader; team leadership behaviours are the responsibility of each team member. Leadership is distributed among the team because all team members take part in problem solving and decision making (Drury-Grogan and O'Dwyer 2013), set out the team's performance expectations and assess if they have met them, and agree on acceptable interaction patterns (for example, in retrospective meetings), synchronise and combine individual team member contributions (with wallboards and software tools), seek and evaluate information that affects team functioning, determine their own roles, and engage as a team in preparatory meetings and feedback sessions.

The second study is by Moe et al. (2010). They drew on Dickinson and McIntyre's (1997) input-throughput-output teamwork model to explore the nature of self-organising agile project teams in a single longitudinal case study of a Scrum team. The chose this theoretical model because it is one of the few theories that explain self-organising teams. This is because it includes concepts of shared leadership and double-loop learning (i.e., reflection and subsequent adjustment) which are characteristics of self-managing agile teams (Cao and Ramesh 2007). This research concluded that the model of Dickinson and McIntyre (1997) explained their findings in many dimensions, but was not able to explain all teamwork aspects and needed to be extended with trust and shared mental models.

The third study, recently published by Yu and Petter (2014), argues that shared mental models theory explains how certain agile practices facilitate the formation of a shared mental model. A shared mental model, also referred to as a team mental model, is a team cognition construct from the organisational behaviour domain. This theory proposes that effective teams develop and maintain a shared understanding of the tasks and relationships within the team to accomplish team goals. Mohammed et al. (2010) surveyed 15 years of research on team mental models and found, "A team mental model (TMM) was introduced as a way to capture the implicit coordination frequently observed in effective teams and to further understand how teams operate in contexts that are complex, dynamic, and uncertain" (p. 876). They defined a team mental model as "team members' shared, organized understanding of knowledge about key elements of the team's relevant environment" (Mohammed et al. 2010, p. 882). Yu and Petter (2014) also argued that practices from Extreme Programming and Scrum such as the system metaphor, stand-up meeting, and on-site customer achieve a team shared mental model in the project team.

### 1.3 Conceptual Framework

The conceptual framework for this paper was adapted from the Big Five theory of teamwork (Salas et al. 2005). As previously discussed, the Big Five theory is highly cited, well-grounded in research literature, and has previously been used in the agile software development context. In addition, Big Five theory encompasses most of the concepts in Dickinson and McIntyre's (1997) teamwork model and incorporates the shared mental model concept discussed Mohammed et al. (2010). Table 1 compares the concepts in these models.

| Concepts | Teamwork theories | | |
|---|---|---|---|
| | The Big Five (Salas et al. 2005) | Dickinson and McIntyre's model (1997) | Team mental model (Mohammed et al. 2010) |
| Adaptability | ✓ | | |
| Backup behaviour | ✓ | ✓ | |
| Closed-loop communication | ✓ | ✓ | |
| Mutual performance monitoring | ✓ | ✓ | |
| Mutual trust | ✓ | | |
| Team leadership | ✓ | ✓ | |
| Team orientation | ✓ | ✓ | |
| Shared mental models | ✓ | | ✓ |
| Coordination | | ✓ | |

*Table 1 Teamwork theories and concepts previously applied to agile software development*

Before applying the Big Five teamwork theory to study teamwork in the agile software development context it was first adapted to better align the concepts in the theory with the agile philosophy. To do this, the Big Five concept of team leadership was relabelled as shared team leadership and a definition was written by combining the original Big Five definition of team leadership with Hollenbeck et al.'s (2012, p. 87) definition of the autonomous team. This led to the following definition.

> *Shared team leadership* is characterised by the team taking responsibility for completion of a variety of tasks, including team maintenance functions. Team members jointly take responsibility for directing and coordinating their activities, assess team performance, assign tasks, develop team knowledge, skills and abilities, motivate team members, plan and organise, and establish a positive atmosphere. The team jointly identifies and solves ill-defined or poorly structured problems, determines performance expectations and acceptable interaction patterns, synchronizes and combines individual team member contributions, seeks and evaluates information affecting team functioning, clarifies team roles, and engages together in preparatory meetings and feedback sessions.

All other definitions for the concepts in the conceptual model used in this paper are as stated in the original Big Five theory. The concept of coordination, proposed in Dickinson and McIntyre's (1997) model, was excluded from this study because it is an outcome of the other teamwork factors. Therefore, the conceptual framework for this study included the teamwork concepts of the Big Five model with the definition of team leadership adapted to become "shared team leadership".

## 2 Research Method

The research question was addressed following guidelines for positivist case study research in information systems research (Dube and Pare 2003; Eisenhardt 1989; Eisenhardt and Graebner 2007; Keutel et al. 2013; Pare 2004; Yin 2003). The unit of analysis, or case, was the agile software development project. To search for evidence for the teamwork concepts, we selected three cases from a single organisation. This meant that some important environmental factors influencing the projects were informally controlled. Each project was influenced by the same organisational culture, all project team members had volunteered to use agile software development reducing the influence of resistance, and all project team members received the same training in agile values and practices from the same trainers, and had the same agile coach and supervisory manager. In addition, all projects were resourced identically, and took place in the same physical location. Furthermore, each project was supported by the same IT infrastructure unit, used the same underlying mainframe system, and complied with the same interface design guidelines and quality expectations.

The organisation was a commercial bank in Wellington, New Zealand. The cases were selected from a pool of nine ongoing and independent projects carried out simultaneously on the same floor of a large open plan office. The senior manager responsible for all of the projects was interviewed as was the agile coach, who was an external consultant. The cases were selected because the agile coach identified these cases as ranging broadly in their complexity (based on technical difficulty and number of stakeholders) and their variation in adopting agile software development practices. The research was restricted to three cases because a single researcher was available to undertake data collection.

The cases were code named Globe, Tech, and Rock. Each project team was creating a high-value business application, and each project was a typical agile software development project with about 10 participants, and was at least 1/3 complete and ongoing at the time of data collection. Table 2 shows details of the cases and interviews. In each project, the team leader was interviewed, and was then asked to nominate further project team members to be interviewed. These team members were selected because they took different roles within the project and agreed to be interviewed. Up to five people from each project were interviewed using a semi-structured interview schedule (available at Strode (2015)). The data collection was restricted to five people to minimise interrupting the work of the project while providing adequate depth of information. Interviews were recorded and transcribed in full. Additional data was collected, whenever it was available and permitted. This included observation and researcher notes taken at selected stand-up meetings and product demonstrations, photographs of Scrum wall boards showing stories, tasks, and task allocations, photographs of Burndown charts, publicly available data from the organisation's web site, and system documentation and project documentation such as organisation charts, interface designs, and example Kanban cards. This non-interview data was used as supporting evidence to assess how agile practices were adopted within the project.

| Case | Globe | Tech | Rock |
|---|---|---|---|
| Organisation level interviews | Senior manager of all agile projects<br>Agile coach for all agile projects (external consultant working on site) | | |
| Roles of interviewees | Business lead/Product owner [EC01]<br>Agile lead/Scrum master [EP01]<br>Business analyst/Agile lead [ET01]<br>Joint interview<br>• Senior analyst programmer [ET02]<br>• Tester [ET03] | Online banking business leader (customer proxy) [FP01]<br>Test analyst [FT01]<br>Mainframe Architect (domain specialist) [FT02]<br>Mainframe developer [FT03]<br>Senior programmer/Agile lead [FT04] | Senior business analyst [GP01]<br>Senior analyst programmer [GT01]<br>Test analyst [GT02]<br>Technical designer [GT03] |
| Team size | 10 | 10 | 6 to 12 varying over the project duration |
| Software product profile | Foreign exchange service | Transaction notification service | Online statements |
| Agile method adopted | Scrum | Scrum | Kanban-Scrum hybrid |
| **Key**: codes such as [EC01] refer to the source of quotes appearing in Tables 3 and 4 | | | |

*Table 2. Project profiles and interview information*

Qualitative content analysis, as described by Schreier (2014) was used to analyse the data. This is a systematic method for describing the meaning of qualitative data and is carried out by assigning parts of the source material to the categories of a coding frame. In this study, we used directed content analysis, which is appropriate when the coding frame is based on existing theory (Hsieh and Shannon 2005). Following Schreier (2014), the adapted Big Five theory was used as a coding frame, a pilot analysis was performed on transcripts from one case to assess the applicability of the framework, the coding was applied by a single coder over two points in time (a mechanism to improve validity), and the framework was then applied to all of the interview data with the aid of the HyperResearch™ tool.

In case study design, Yin (2003) recommends triangulation to improve the validity of the study by corroborating the findings. Data triangulation is the use of different data sources to provide evidence on which to base findings. This was achieved in two ways. Firstly, data was collected on the same topic from different sources of the same type. There were three data sources: three cases of agile software development. Then within each case, there were four or five data sources because up to five people in each project were interviewed. The semi-structured interview technique followed a pre-designed schedule including closed and open-ended questions allowing for further probing when necessary. Secondly, data was collected from different types of data source as mentioned in the previous section. Reliability was achieved using a transparent research process so that the findings can be traced from initial theory to final conclusions (Yin 2003). This study had an existing conceptual framework that was applied in the same manner to each case to identify instances of evidence for teamwork concepts. To illustrate the type of evidence found in support of each factor, Tables 3 and 4 show a single example from each case for each factor. Overall, multiple instances for each teamwork factor were found in each case except where noted in the tables.

## 3 Findings

Senior management decided that the bank needed to be more flexible in adapting to changing market conditions and this motivated the adoption of agile software development. All senior management and the project management office were trained in the agile philosophy and agile project management, and agile practices were introduced to support the initiative at the operational level. Within the IT section, the adoption was viewed as a way to improve delivery times and system quality. The manager said there was "a general perception across the organisation that we have a fair number of project failures using the waterfall methodology", and "we always took on really large scale projects and they [would be] delivered over a year and a half to two years, and during that time we would just get nothing out the door for our customers".

The bank adopted agile software development in selected projects that were concerned with improving customer self-service. A developer explained, "we were told that agile was going to be [used], there was going to be a whole lot of pieces of work [done using agile], they did a big restructure, and… if you were interested in agile put your hands up." The bank renovated a floor of their building to provide space suitable for multiple agile teams. A Scrum master said, "A decision was made that [the Bank] was going to invest in agile as a methodology, and so we all had some training…And we were given coaching…, they hired some people in to specifically spend time with us." A consultant agile coach introduced agile values and practices to the project teams in training sessions and provided coaching in the initial weeks of each project. Agile practices were progressively introduced into each project, and the coach regularly assessed the uptake of practices by each team using a proprietary assessment measure. Each project team chose to adapt agile differently.

### 3.1 Project Globe

Globe was the most straightforward project. The Business Lead (acting as a Scrum product owner) captured requirements by consulting weekly with four stakeholders located in Auckland, and three further interested stakeholders were kept informed of progress. The Business Lead worked three days in Wellington with the team and three days in Auckland with the stakeholders. User stories came from those stakeholders, the Lead's experience in solving customer's problems, and from the project team.

A notable issue was explained by the Agile Lead (Scrum master), "None of the operational support people are in this building…some of the [mainframe and network] support is done in Australia. Some of that is done by teams at the [bank] in Wellington, some of it's done by [a telco] or people in Australia…Anything that you have to get done by people in Australia always takes much longer and requires far more paper work. They're not at all agile in the way they operate. It's very bureaucratic. And getting people to do work outside our immediate team is always more time consuming because they're not part of our agile team, they generally have lots of projects they're doing work for, and even if our work is a priority for them it has to be fitted in around all their other day to day work." There were similar issues with external parties, "The project that we're working on is very design driven. It's an internet banking project,…but our design team is very small so we've got some people from [X]…an external company, to come in and do some of the more specialised areas… and of course, they're not always available at the drop of a hat when we want them." Another issue was the team makeup. The Business Lead explained "we've got five developers and three testers. So we're a bit developer heavy. And so that can provide challenges with getting the work done properly. Because ideally you have the stories and a script prioritised and you finish one story then you start the next and then you start the next. But when you've got five developers, they can't all work on one story easily, so it's very difficult to try and keep to those agile principles of doing things in a prioritised order."

This project adopted the Scrum process with two week sprints, daily stand-ups, and co-location with line of sight to their Scrum wallboard. They used sprint planning meetings, product demonstrations to stakeholders, and retrospective meetings within each sprint. The Business Lead attributed their success in adopting agile practices to the familiarity of the team, "a lot of us have worked together; probably 80% of the team have worked together for seven years…It's made it much easier for us to adopt Agile, because we already knew each other quite well…got on well and worked together well." A developer also thought existing team characteristics were more important than agile practices; he said, "I don't think Agile helps you be positive. If you are positive then Agile works better."

Globe provided evidence for each of the Big Five teamwork factors (as shown in Tables 3 and 4): adaptability, backup behaviour, closed-loop communication, mutual performance monitoring, mutual trust, shared mental models, team leadership, and team orientation.

| | Globe | Tech | Rock |
|---|---|---|---|
| Adaptability | **Team agrees to adapt**<br>"When we were first doing our Agile Sprints, we would do … spec workshops, which is where you go through the stories to understand the business intent of upcoming stories…we used to…have these meetings once a week for a couple of hours…But…we found that when we got to the Sprint and we decide what we've going to do, sometimes the Business Lead will have changed her mind on what her priorities were, or sometimes you couldn't do what we thought we were going to do. And… all that work had been wasted. So we decided that rather than do the… workshop…two weeks in advance, we would do it immediately before our Sprint planning. And that's worked out better." [EP01] | **Team agrees to adapt**<br>"So we failed…when we went to show it the test environment wasn't up and running. So we couldn't actually demo it…we just had to talk about it… everyone in the team was very upset about that because it … looks bad … So …one of the decisions we made was that for the show and tell we're going to move it to a different day, to a Tuesday, and rearrange all those other meetings slightly to make it work. And … we were going to put one person in charge of the show and tell and we'd rotate that. And…that person was…responsible for checking things and chasing up if the environments were down,…making sure the rooms are booked, and the video conferencing is working." [FP01] | **Lack of adaptability**<br>The project team found it difficult to adopt agile practices. "Before this project, I was already involved in another Agile project where we had… Scrum… – but that was also more of a hybrid, because we started off with requirements that were meant to be from Waterfall sense of development… But we ended up making it an Agile project. But the difference with that one and this one is that this one, there we had the planning sessions … we had stories, and we had estimations for the stories, and we kept to the how many hours you have within the Sprint, and then we did the actual Sprints and we did the actual show and tells. So I had, in comparison to this one, it had more Agile components in it." [GT01] |
| Backup behaviour | **The wall with stories supports backup behavior**<br>"And of course, there's a bit of cross work. Sometimes if there's a lot of testing work outstanding, the developers will help with some testing. And there are some tasks [on the wallboard] that everyone on the team will work on, depending on who is free and who is available." [EP01] | **The wall with stories supports backup behaviour**<br>"When we are say, finished with a task and there's nothing more to do, then we'll go and ask if there's a sequence of things that one person is intending to do, if there's some backlog or we ask the person if he is alright or if it's alright for us to jump into one of the tasks." [FT03] | **The wall with Kanban cards supports backup behaviour**<br>"And some of it was at the wall…, 'Okay when are you going to be able to finish all that testing?' and I would say, 'Well I've got this and this and this to do,' and Madhup would say, 'Well I'm just about finished this and I haven't got anything coming for a week so I can do something.'" [GT02] |
| Closed-loop communication | **Communication for helping supported by proximity**<br>"Because we help each other, we communicate a lot and we always can jump in and say, "Oh can you please have a quick look here for me," or approach [Carlos], like this morning and say, "I think I've got it right. Can you just come and have a look to make sure that I've got it right?" And he just comes and look over my shoulder and just says, "Yep, yep, yep."[ET02] | **Communication for achieving understanding supported by proximity**<br>"I send communications to… the developers as a whole. I'll also talk to the host developers…or else I'll go and talk to…the BA. There might be something I don't fully understand in the acceptance criteria. Or the…business lead if I…have a question there. I receive communications from everyone on the team… I communicate with everyone on the team and I receive communications from everyone on the team." [FP01] | **Communication for problem solving supported by proximity**<br>"If I had an issue or a problem then I would go straight to, he was just sitting right in front of me, [Valor]. And [Kate] was…on my left….So we did have those conversations …the face to face always helps. We were able to explain right away and the turnaround was quick. …. The tester was beside me as well. So you had the test analyst, you had the three components people…right beside each other." [GT01] |
| Mutual performance monitoring | **The wall with stories supports monitoring**<br>"… we have learnt from mistakes with sprints … say developers working on stories for 8 days out of 10 and then deliver everything to us in the last two days and we would have massive testing tasks in a short time so that was like a mini Waterfall within the sprint and that's not how it is meant to happen. We are meant to finish the first story then move on to the next one and so on or at least start a couple of stories but finish them before we move towards the bottom of the wall. So we keep an eye on each other and if we see that we tend to [take action]" [ET02] | **Co-location supports monitoring**<br>"So the communication within the team I think is fundamental. And you see it when someone forgets to tell someone something, that it interrupts what they were actively doing, and they go "what's just gone on here?" And quickly do a run around the team and find out someone's done something that's impacted another person. So, communication is definitely the biggest one. That doesn't mean that it's perfect … you still get small break downs that cause interruptions." [FT04] | **No evidence in the transcripts** |

*Table 3 Case evidence for teamwork factors*

|  | Globe | Tech | Rock |
|---|---|---|---|
| Mutual trust | **Trust that the team does their best**<br>"But we have such a good team that everybody is just communicating…and never says something bad about another person, we generally believe that everybody does his best to develop something … [ET02] Best job to their abilities and …we don't really question people's abilities. Everyone does their best."[ ET03] | **Trust that the team provides accurate information**<br>"So the negotiation in our team is possibly raising technical stories that have little or no business benefit, even though it directly contributes to the end product, and is required. And you negotiate with the business lead…I talk to [her] about it, obviously [she] trusts the development people to give her accurate information about why it's important and it's very easy and simple conversation." [FT04] | **Trust that the team has suitable skills**<br>"The team dynamic. The core team members all understood, once we'd got into the project, what was needed. The people doing all the coding were all on the same page and all competent people with good skills." [GT02] |
| Shared mental models | **Specification meetings support a shared mental model**<br>"I find the spec workshop, although I find it boring as hell sometimes, it's interesting, everybody's getting an understanding of what's supposed to be happening. I don't necessarily understand what they're all talking about, but there's clarity…between everybody. So that nobody, theoretically, is thinking when I say "I want to change a rule in the host system" that I'm saying I want to change a rule somewhere else. Because we've talked about it, everybody has an understanding of what I want." [EC01] | **Planning and stand-up meetings support a shared mental models**<br>"We've gone back to our desks, and we're gearing up for the real start of the coding, and the developing, and it just occurred to us that we just said it's better for you to do these, it's better for me to do these, and you and you and you. So it's like a Gentleman's Agreement that the pieces of work are grouped in. So in our minds, is not set or written, or anything, but what we have is in our inner mind we already kind of talk about this piece of work is more suited for you." [FT03] | **Lack of shared mental model**<br>"Another thing that we probably could have done better, is had a bit more of an overview of … what we're trying to do. But [new team members are] sort of thrust in and said "do this" with no context…I know for some of the other host developers that came on, it would have probably been really good for me to sit down with them and talk about the project and give them a bit more context about what they're actually doing, and why." [GP01]<br>**Stand-up meetings support shared mental model**<br>"So the stand-up meeting I think contributed a lot to it. Because you'd know during those meetings what people were working on, and who was working on what." [GT01] |
| Shared team leadership | **Strong shared team leadership**<br>"And it's …, not peer pressure as such, but it's almost team responsibility and someone has to do it … normally what happens is someone will say, "Oh yeah I can do that." They will delegate themselves." [ET01] | **Strong shared team leadership**<br>"Well, it's not assigned, you take it, …we'll have a bunch of stories that are mainframe centred, and a bunch of tasks that are all related to those stories, and obviously there's got to be some order in the tasks, because you can't test before you build. But nevertheless, we'll decide just between ourselves, and we might say "Nick, you take that one, Livia you take that, and I'll do this one". [FT02] | **Weak shared team leadership**<br>"The Kanban lead … if he wasn't there for a couple of days then … sometimes things would just get left without being attended [to]. Like if something was supposed to be sent across to [the archiving company] but he wasn't there to do it then it would just sit…Anybody could have done that [task]… you know, you need somebody there making sure that everything is still ticking along." [GT02] |
| Team orientation | **Team-oriented commitment to the task**<br>"The Business Lead will want ten stories done and the project team will only want to do five and there'll be a bit of to-ing and fro-ing and they might agree to six, or they might agree to seven, or they might only want to do five. But it's up to the team. The Business Lead will always try and put a bit of pressure on, which is her job. But it's up to the team to commit to something." [EP01] | **Individual commitment to the task**<br>"But there's a sense of ownership, so if you pick something up [a task] you tend to see it through. And again, it's still part of the culture that was here beforehand anyway. The culture I thought was very good to begin with. So switching to the Agile in terms of the culture itself is not too different apart from the fact that you've just unsettled everyone." [FT04] | **No evidence in transcripts** |

*Table 4 Case evidence for teamwork factors continued*

## 3.2　Project Tech

This project was assessed by the Agile Coach as particularly technically challenging, which caused the project team to struggle with the adoption of agile practices. This project had a Business Lead, an Agile Lead who was also a developer (Scrum master), six additional developers, and two testers. The technical environment meant that the team included mainframe developers, Java developers, and testers. These groups formed subgroups within the team. Other people, such as interface developers and database analysts were not part of the team and their services had to be requested. The stakeholders in this project consisted of two senior managers in the bank. The lack of direct contact with the customer was noted by team members as a problem affecting their motivation because they did not get direct feedback on the quality of their work, and they did not have someone to assist with prioritising their stories.

This project team adopted the Scrum method. Stories for each sprint were selected by the Business Lead and one domain expert from a business requirements document (the team called this 'the spec'). This document was developed before the project team began using the agile approach. This project team used 3-week sprints, spec workshops for identifying stories from the spec document, sprint planning meetings with story estimation, daily stand-ups, a scrum wallboard, user stories and tasks. 'Show and tell' sessions (product demonstrations) were held regularly to show the various stakeholders the current software version and ask for feedback. Some stakeholders used videoconferencing to attend and all team members attended. The team also used test-driven development and regular system integration.

Tech provided evidence for each of the Big Five teamwork factors (as shown in Tables 3 and 4): adaptability, backup behaviour, closed-loop communication, mutual performance monitoring, mutual trust, shared mental models, team leadership, and team orientation.

## 3.3　Project Rock

Rock was a complex project because of its multiple stakeholders and need to integrate with multiple systems within the bank. Internally, there were seven stakeholders, each one was the head of a division. There were four external stakeholders, each one an individual company (an airline, a credit card company, an archiving company, and a mobile alert company). The new system had to integrate with each of the external stakeholders' systems as well as change the bank's mainframe system, provide a new interface, and web services. In addition, the project had started years before but was not well-supported as a single business analyst had spent some months working alone to develop use case models. The project slowly became populated with a team, then the use case models were used to develop Kanban work items. The project had a fluctuating team size. At the time of interviews seven core full-time employees including a project manager (assigned to multiple projects), a business analyst, a mainframe specialist, two testers, and two Java developers. As the project progressed part-time staff were added including a Java developer, three mainframe developers, and a middleware specialist. A number of problems were reported by the interviewees. They had to wait and switch tasks because of the slow response of some external stakeholders, they spent a lot of time negotiating the new system design with the stakeholders, new staff on the project had to spend time becoming familiar with the project complexities, and the tester's work was complicated by the bank's test databank, which was shared with other projects.

Rock adopted a hybrid method that included Kanban and elements of Scrum. They used a Kanban wallboard without sprints, held stand-up meetings at their Kanban board, held two retrospectives during the project, and did not have product demonstrations. The project team seemed unsure about their agile adoption, although the project was labelled as agile by the bank. Some members of the project team seemed confused about this. The business analyst described the method as "waterfall-on-a-wall", and a tester described it as "…waterfall using a wall, and stand-ups." The test analyst said, "I would say the business analyst was also the sort of the lead, started off calling him an Agile lead but because we weren't Agile we tried to shy away from that phrase, probably call him the Kanban lead".

In the Rock team, there was little cross-over between team roles (this also occurred in Tech); fairly strict role division was maintained between Java developers, middleware developers, mainframe developers, and testers. People who came into the project maintained their specialism and did not move outside their role. Some of the project team had worked for the bank for a many years, but others were new to the bank.

Rock provided evidence for only some of the Big Five teamwork concepts (see Tables 3 and 4). Backup behaviour, closed-loop communication, and mutual trust were all evident. There was evidence for a lack of shared team leadership, lack of shared mental models of the team and the task, and lack of adaptability. There was no evidence for mutual performance monitoring or team orientation. There could be many reasons for this lack of fit. The team did not adopt a full Scrum package of practices, the team was large, there were multiple stakeholders for the project, and team members were added to the project slowly over time, especially during times when the workload rose. Any one of these factors, or all of them in combination, could account for the lack of theoretical fit. Large teams with no distinct formation period could find it difficult to form a shared mental model, they are less likely to be adaptive and share team leadership (become self-managing) since new team members might not challenge existing team protocols when they enter the team.

## 4   Discussion

This series of case studies confirms that Big Five teamwork theory applies to agile software development, but only when the team leadership concept is redefined as *shared* team leadership. Furthermore, for these independent cases of agile software development, the adapted Big Five teamwork theory fit only two of the three cases. We found evidence that adopting agile practices does not guarantee that each of the teamwork concepts in the model will occur. This was illustrated in the case of project Rock, which adapted a hybrid agile method and did not provide evidence or provided negative examples, for shared team leadership, shared mental models, adaptability, mutual performance monitoring, and team orientation. This lack of theoretical fit could be attributed to the number and type of agile practices adopted or could be due to other contextual factors such as team formation, team size, or project complexity. These factors should be investigated in further agile teamwork research.

The findings provided evidence that certain agile practices adopted by the project teams directly support some of the Big Five teamwork concepts. Evidence, in Tables 3 and 4, indicates that the wallboard displaying stories and tasks, supports backup behaviour and mutual performance monitoring. The wallboard displaying Kanban work items supports backup behaviour. Communication is supported by proximity and contributes to helping behaviour, achieving understanding, and problem solving. Co-location supports mutual performance monitoring. Shared mental models are created during team meetings including specification meetings (not a typical Scrum practice but peculiar to the bank context), Scrum planning meetings, and daily stand-up meetings.

This research makes theoretical and practical contributions. The theoretical contribution is a teamwork theory that applies to agile software development projects. This research has shown that the Big Five teamwork theory proposed by Salas et al. (2005) is not universally applicable and needs adjusting for the agile software development context where teams act as self-organising groups and share team leadership among team members. Thus this study confirms the finding of Moe and Dingsøyr (2008). This research contributes to practice by providing evidence that teamwork is directly supported by colocation, having publically viewable wallboards showing stories and tasks or work items, and holding regular structured meetings to share information and reflect on and adjust the team process.

This research is limited because the findings are based on data from three cases. More cases or more comprehensive data collection encompassing more interviews of all project team members, a greater a level of participant observation, or longitudinal cases might have drawn forth more evidence for some of the concepts in the teamwork theory. In addition, case study research is not generalizable to populations, so the findings are not applicable to all agile software development projects.

## 5   Conclusion

This research built on prior teamwork studies to propose an adapted Big Five teamwork theory for agile software development projects. Empirical data from three cases showed that the theory is applicable in some cases, but is not universally applicable. This could be due to the agile practices adopted or other factors. As this study has provided confirmation that an adapted Big Five theory of teamwork applies to some agile software development projects, future work should study 1) the influence of high functioning teams on agile software development adoption, and 2) the causal relationship between the use of agile practices and effective self-organising teamwork.

## Acknowledgements

We are grateful to the bank management, the consultant agile coach, and the case study participants for readily giving their time and insights to support this research.